\documentclass{aa}  

\usepackage{graphicx}
\usepackage{gensymb}
\usepackage[T1]{fontenc}
\usepackage{amssymb}
\usepackage{natbib}
\usepackage{tabularx}
\usepackage[utf8]{inputenc}
\usepackage[english]{babel}
\usepackage{txfonts}
\usepackage{ulem}
\usepackage{xcolor}
\usepackage{hyperref}
\begin{document}

\title{First ALMA observations of the HD 105211 debris disc: A warm dust component close to a gigayear-old star}

\author{
Qiancheng~Yang\inst{\ref{gzu}}
\and
Qiong~Liu\inst{\ref{gzu},\ref{cam}}
\and
Grant M. Kennedy\inst{\ref{war}}
\and
Mark C. Wyatt\inst{\ref{cam}}
\and
Sarah Dodson-Robinson\inst{\ref{del}}
\and
Rachel Akeson\inst{\ref{cal}}
\and
Nenghui~Liao\inst{\ref{gzu}}
}

\institute{
Department of Physics, Guizhou University, Guiyang 550025, China\label{gzu}\\
\email{qliu1@gzu.edu.cn}
\and
Institute of Astronomy, University of Cambridge, Madingley Road, Cambridge CB3 0HA, UK\label{cam}
\and
Department of Physics and Centre for Exoplanets and Habitability, University of Warwick, Gibbet Hill Road, Coventry CV4 7AL, UK\label{war}
\and
Department of Physics and Astronomy, Bartol Research Institute, University of Delaware, 217 Sharp Lab, Newark DE 19716, USA\label{del}
\and
Caltech-IPAC, Pasadena, CA 91125, USA\label{cal}
}
\date{Received XXX; accepted YYY}

\abstract
	{Most debris discs consist of a gas-poor, cold dust belt located tens to hundreds of astronomical units away from the host star. Many cold dust belts exhibit distinct structures attributed to the dynamic interaction of planetary systems. Moreover, in a few systems, additional warm components can be found closer to the central star, resembling the asteroid belt or zodiacal dust in our Solar System.
}
	{In this work, we investigate the structure of the disc surrounding the nearby F2V star HD 105211 ($\eta$ Cru, HIP 59072), which has a warm excess (seen with \textit{Spitzer}/MIPS at 24 $\mu$m) and a potential asymmetry in the cold belt (seen in the \textit{Herschel}/PACS images).
}
	{ We applied the CASA pipeline to obtain the ALMA 1.3 mm continuum images of HD105211. Then we constructed the spectral energy distribution (SED) of the system and performed Markov Chain Monte Carlo (MCMC) simulations to fit a model to the ALMA visibility data. To characterise the disc asymmetry, we analysed the ALMA images of two individual observation blocks (OB1, conducted on 2017 March 28, and OB2, conducted on 2018 May 9) and compared them to the previous \textit{Herschel} images.
}
	{Our modelling reveals that the disc around HD105211 is a narrow ring (23.6$\pm$4.6 au) with low eccentricity ($e\leq0.03$) positioned at a distance of 133.7$\pm$1.6 au from the central star, which differs from the broad disc ($100\pm20$ au) starting at an inner edge of $87\pm2.5$ au, inferred from the \textit{Herschel} images. We found that both observation blocks show excess emission at the stellar position ($>3\sigma$), while OB1 shows an offset between the star and the phase centre ($\sim0.3\arcsec$), and OB2 shows brightness clumps ($\sim2\sigma$). We used a two-temperature model to fit the infrared SED and used the ALMA detection to constrain the warm component to a nearly pure blackbody model.
}
	{The relatively low ratio of actual radius to blackbody radius of the HD105211 debris disc indicates that this system is depleted in small grains, which could indicate that it is dynamically cold.
	The excess emission from the stellar position suggests that there should be a warm millimetre-sized dust component close to the star, for which we suggest two possible origins: in situ asteroid belt or comet delivery.
}

\keywords{
	methods: observational 
        --
	stars: circumstellar matter 
	-- 
	infrared: planetary systems
	-- 
	stars: individual: HD 105211
	}
 \titlerunning{First ALMA observations of HD 105211}
\authorrunning{Yang et al.}
\maketitle

%________________________________________________________________

\section{Introduction}
\label{sec:Introduction}

Debris discs are dusty ring-like structures that surround stars and contain exocomets, asteroids, and dust, and they are analogous to the asteroid belt and the Kuiper Belt in our Solar System \citep{1993ARA&A..31..129L}. Representing the final stage of circumstellar disc evolution, debris discs hold valuable information about the evolution and dynamical history of a system. Interactions of planetary components lead to destructive collisions of larger planetesimals, known as collisional cascades \citep{2008ARA&A..46..339W}, and cause the solid components of debris discs to have a wide radius distribution, from small asteroid-sized objects to micrometre-sized particles. Optical and infrared light from the star in the debris disc is absorbed by the solid components and then reemitted at longer wavelengths. At infrared and millimetre wavelengths, the thermal emission from dust can be much brighter than the central star.

The spectral energy distribution (SED) of a system probes the stellar and disc properties. Multi-wavelength photometry provides valuable information such as an estimate of the dust temperature and its radial location \citep{2007ApJ...658..569W, 2011MNRAS.414.2486K, 2014AJ....148....3L}. Surveys have shown that most debris discs are located at distances of tens to hundreds of astronomical units from the star they surround, similar to the Kuiper Belt. At such distances, the dust have a low-temperature and are referred to as "cold dust" \citep[e.g., ][]{2013A&A...555A..11E, 2014ApJS..211...25C, 2018MNRAS.475.3046S} However, a small fraction of debris discs also contain warm or even hot dust, which is inferred to be closer to the star, at distances of just a few astronomical units or less, similar to the asteroid belt or zodiacal dust \citep{2013A&A...555A.104A, 2017ApJ...845..120B, 2021RAA....21...60L}. This hot dust is mainly found in debris discs with ages younger than 100 Myr \citep{2013MNRAS.433.2334K}, but it has also been observed around some older debris discs, such as $\eta$ Corvi \citep{2017MNRAS.465.2595M}.

Further insight into the physics of debris discs has been provided by resolved images at multiple wavelengths that trace grains of different sizes and reveal diverse morphologies. Optical and infrared observations trace micron-sized grains, which are influenced by factors such as radiation pressure \citep{2009ApJ...705..314S}; Poynting-Robertson drag \citep{1979Icar...40....1B}; and even the interstellar medium \citep{2014A&A...569A..29M}. Such observations have unveiled specific features such as extensions \citep[e.g., HD 15115,][]{2014AJ....148...59S}; warps \citep[e.g., $\beta$ Pic,][]{2000ApJ...539..435H}; and wings \citep[e.g., HD 32297 and HD 61005,][]{2014AJ....148...59S}. Millimetre and submillimetre observations trace larger millimetre-sized grains that are less affected by non-gravitational forces. As a consequence, (sub)millimetre observations provide insights into the location of the planetesimal belt, where parent body collisions occur, which makes them especially critical to understanding the physical mechanisms shaping debris discs. There are also various structures observed in (sub)millimetre images, including rings \citep[e.g., HD170773; HD181327; HR 4796A,][]{2019ApJ...881...84S, 2016MNRAS.460.2933M, 2018MNRAS.475.4924K}; wide discs influenced by multiple planets \citep[e.g., HR 8799,][]{2018ApJ...855...56W} or other effects \citep[e.g., 49 Ceti;][]{2017ApJ...839...86H}; gaps potentially caused by orbiting planets \citep[e.g., HD 107146; HD 92945,][]{2018MNRAS.479.5423M, 2019MNRAS.484.1257M}; eccentric rings that may be sculpted by eccentric planets \citep[e.g., Fomalhaut; HD 53143,][]{2017ApJ...842....8M, 2022ApJ...933L...1M}; and clumps that may result from transient collisions or planetary resonances \citep[e.g., $\epsilon$ Eri,][]{2017MNRAS.470.3606H}. All of these structures reflect the dynamical evolutionary history of debris discs, but their relationship to the early protoplanetary phase evolution remains unclear \citep{2019MNRAS.486..453L}.

A nearby F2V star, HD105211 ($\eta$ Cru, HIP 59072) has an estimated age of 1.34-1.92 Gyr \citep{2022yCat.1355....0G} and is located at a distance of 19.76 pc \citep{2007A&A...474..653V}. We note that it is a spectroscopic binary star, and the companion component remains unresolved in the optical \citep{2012A&A...542A.116A}. The infrared excess of HD105211 was first reported in \cite{2006ApJ...652.1674B} as a bright extended source observed in the Multiband Imaging Photometer for \textit{Spitzer} (MIPS) 70 $\mu m$ image. \cite{2014ApJS..211...25C} presented a two-temperature model in their SED fitting, while \cite{2016ApJ...833..183D} found a significant warm excess at 24 $\mu m$ of \textit{Spitzer}/MIPS based on the \cite{2011AJ....141...11D} model photosphere and their re-reduction of the MIPS data. They also first reported the excess at 70, 100, and 160 $\mu m$ of the Photodetector Array Camera and Spectrograph (PACS) of the Herschel Space Observatory and presented the resolved images. After that, \cite{2017MNRAS.468.4725H} combined these PACS data with photometry from optical to far-infrared wavelengths to better determine the disc architecture. Their simultaneous fit to the Herschel images and the SED suggested a broad disc (100 $\pm$ 20 au) starting at an inner edge of 87 $\pm$ 2.5 au, and the resolved images showed a potential asymmetry such that the NE ansae is closer to the stellar position (86.0 $\pm$ 5.7 au) than the SW counterpart (116.0 $\pm$ 8.2 au). Moreover, evidence from the SED indicates the presence of warm dust close to the star. However, there was no high-resolution imaging to characterise the warm dust at the time.

In this work, we study the disc structure of HD105211 through Atacama Large Millimeter/submillimeter Array (ALMA) 1.3 mm continuum observations. In Sect.~\ref{sec:Observations}, we describe the observation setup and analyse the continuum image and its level of asymmetry. Then, in Sect.~\ref{sec:Model}, we present both the SED fitting and visibility model fitting of the ALMA data to constrain the spatial distribution and properties of the dust. Our fit sets an upper limit on the eccentricity of the disc and identifies a flux excess from the stellar position. In Sect.~\ref{sec:discussion}, we compare the ALMA images with previous images from \textit{Herschel} and discuss the collisional status of the disc as well as the warm inner dust component. Finally, Sect.~\ref{sec:Conclusions} contains our conclusions.

\section{Observations}
\label{sec:Observations}

\subsection{Observational information}

The star HD 105211 was observed with ALMA Band 6 (1.3 mm) under the Cycle 4 project 2016.1.00637.S (PI: Dodson-Robinson, Sarah). 
The observation block was executed in configuration C43-1 with a minimum baseline of 15.1 m and a maximum baseline of 455.6 m. Other observation parameters are listed in Table~\ref{tab:observation} and include the total observation time on source $t_{sci}$, J2000 coordinates (RA and Dec), precipitable water vapour (PWV; a measure of the weather quality), and the beam size. 

We note that the two observation blocks of HD105211 are separated by approximately one year, with different baseline combinations. The existence of two epochs helps in identifying any time evolution in the HD 105211 system whilst confirming the authenticity of each signal. Hereafter, we refer to the observation block conducted on 2017 March 28 as OB1 and the one from 2018 May 9 as OB2.

The correlator was set up to study the dust continuum emission with four spectral windows centred at 223 GHz, 225 GHz, 239 GHz, and 241 GHz; a bandwidth of 2 GHz; and a channel width of 15.625 MHz. In combination, the four windows provided a total bandwidth of 8.0 GHz to study the target emission. 

The data were calibrated using the standard ALMA pipeline in \texttt{CASA5.1.1} \citep{2007ASPC..376..127M}. To reduce the data volume, the calibrated visibilities were time-averaged in intervals of 10s. We carried out continuum imaging and deconvolution using the "tclean" task in CASA. To obtain the greatest signal-to-noise ratio, we reconstructed the image using natural weighing.

\begin{table}
	\centering
	\caption{Summary of ALMA observations at 1.3 mm (Band 6).}
	\label{tab:observation}
	{
		\begin{tabular}{lcc} 
			\hline
			Parameter & 2017 Mar. 28 & 2018 May 09 \\
			\hline
			No. antennas & 44 & 45 \\
			Array & 12m & 12m \\
			$t_{sci}$(min) & 30.17 & 30.11 \\
			J2000 R.A. & 12h06m52.99s & 12h06m52.99s\\
			J2000 Decl. & -64\degr36\arcmin50.11\arcsec & -64\degr36\arcmin50.13\arcsec \\
			PWV (mm) & 2.73 & 1.17 \\
			Baseline (m) & 15.1-160.7 & 15.0-455.6 \\
			Beam size (\arcsec) & 2.25 $\times$ 1.85& 1.41 $\times$ 1.11 \\
			Beam PA (\degr)& 49.6 & -72.6\\
			\hline
		\end{tabular}
	}
\end{table}

\subsection{Results}
\label{sec:ob Results}

\begin{figure}
	\includegraphics[width=0.45\textwidth]{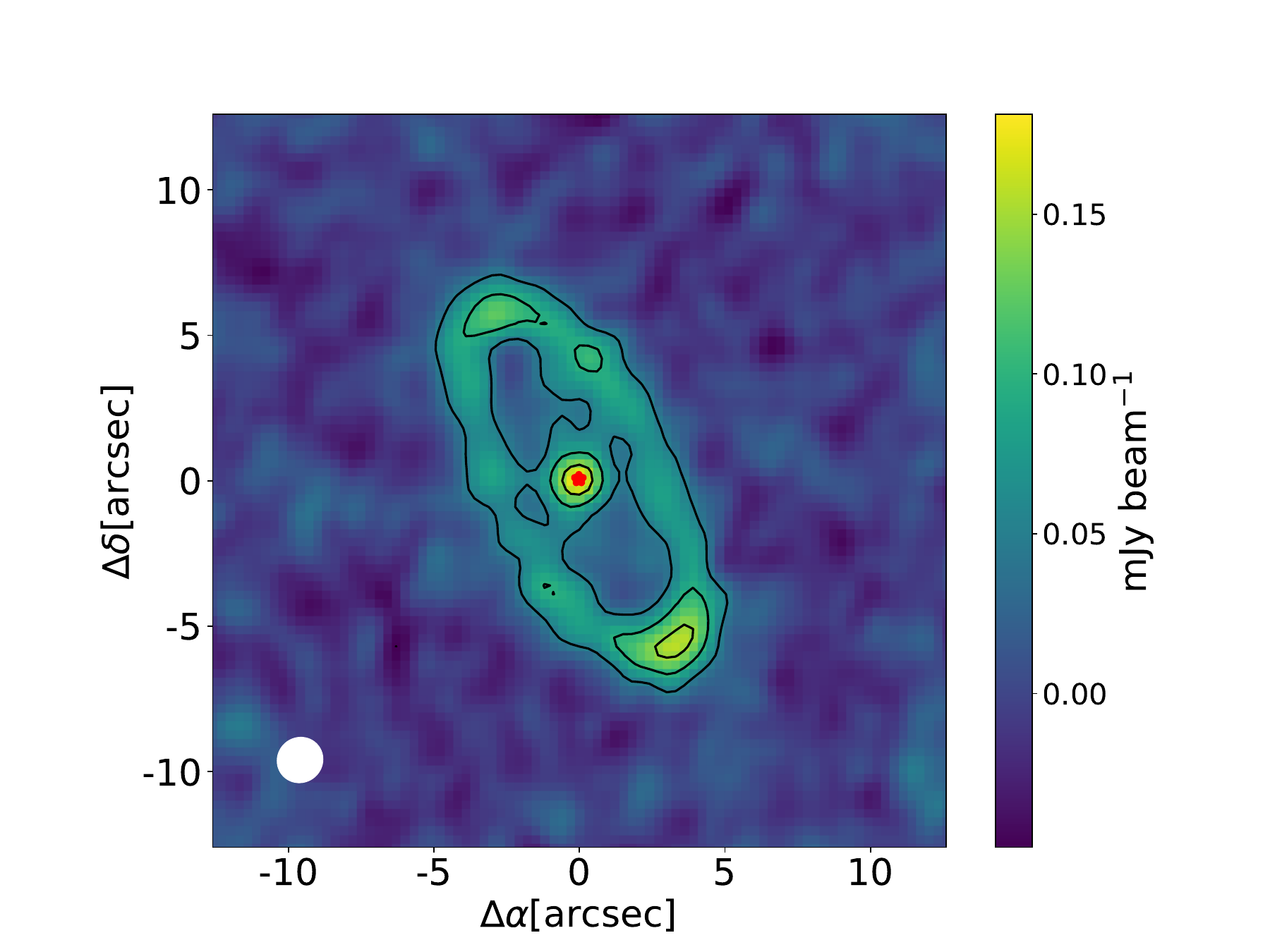}
	\caption{ALMA 1.3 mm continuum image of HD 105211 obtained using natural weights. Contours are drawn at the levels $[-3,3,6,9] \times \sigma_{rms}(rms=16$ $\mu$ Jy $\rm beam^{-1}$). The beam size is represented by a white ellipse ($1.62\arcsec \times 1.57\arcsec$) in the bottom-left corner. The star location is marked as a red star symbol. }
	\label{fig:cont}
\end{figure}

Figure~\ref{fig:cont} presents the ALMA 1.3 mm continuum image of the HD 105211 system. The image was synthesised by combining the visibilities from both observation blocks using natural weighting. The synthesised beam has dimensions of $1.62\arcsec \times 1.57\arcsec$ (32 au $\times$ 31 au) and a position angle of $40.2\degree$as well as an rms sensitivity of 16 $\mu$Jy $\rm beam^{-1}$. A bright point source is prominently featured at the centre, with a peak S/N of 10.5 and a flux of approximately 168 $\mu$Jy $\rm beam^{-1}$. The emission of the central star ($\sim$100 $\mu$Jy, estimated from the SED; see Sect.~\ref{sec:SED}) dominates the point source, but there is also a weaker extended structure that may be attributed to the warm debris disc near the star. The brightness peak of the cold disc is located at a distance of 6.6\arcsec (130 au) from the star. The width of the disc is 1.6\arcsec (32 au), which is close to the beam size, indicating that the full width of the disc is even narrower.

\begin{figure*}
	\centering
	\includegraphics[width=\textwidth]{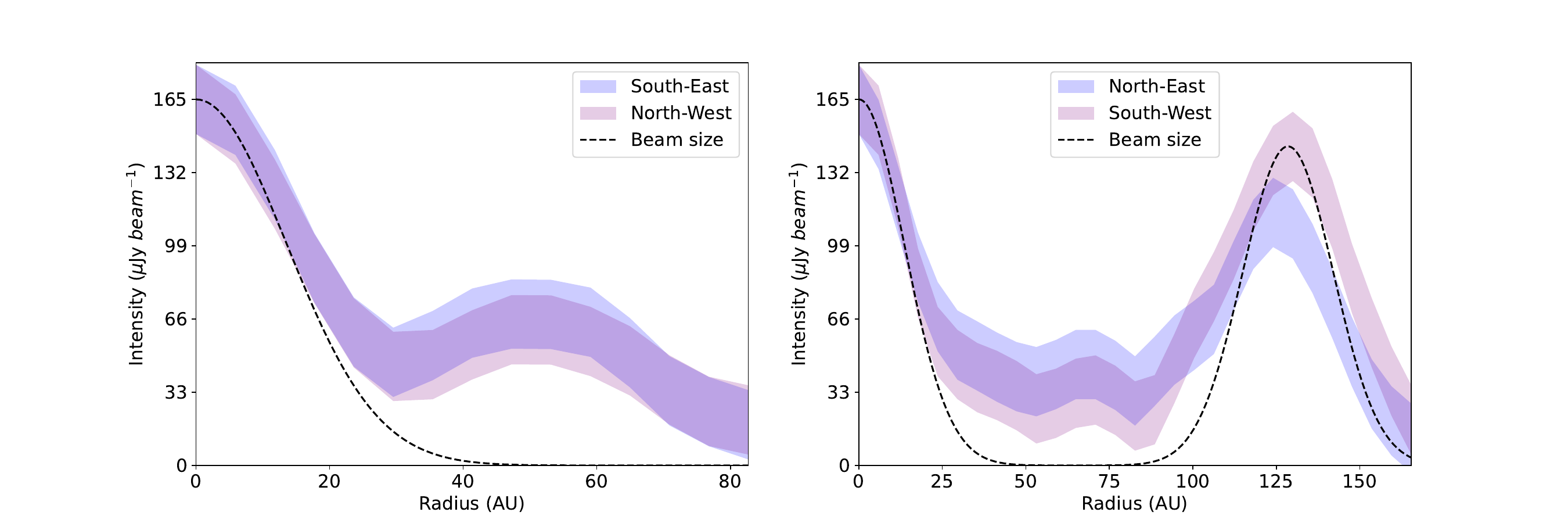}
	\caption{Surface brightness profile of disc cut along the minor (left) and major (right) axis. The blue and purple shaded regions correspond to the 1 $\sigma$ uncertainties of flux, and the dashed line corresponds to the beam size.}
	\label{fig:profile}
\end{figure*}

To further study the disc structure and properties, we analysed the surface brightness profiles along both the major and minor axes with radial averaging using $10\deg$-wide swaths. Figure~\ref{fig:profile} provides the radial profiles along the major axis (left panel) and minor axis (right panel), including the $1\sigma$ uncertainties using the $\sigma_{rms}$ of Fig.~\ref{fig:cont}. To facilitate analysis of the resolved structure, we have included the clean beam in the figure.

The profiles of both the minor and major axes show that the central point source extends beyond the boundaries of the beam. Whether these emissions come from an inner component or just noise needs to be investigated further. Within the minor axis profile, the southeast and northwest directions exhibit similar peak positions and flux densities, indicating the symmetry in these regions. However, the major profile presents a different picture, with the peak flux of SW ($158 \pm 16$ $\mu$Jy $\rm beam^{-1}$) exceeding that of the opposite {\it ansa}($125 \pm 16$ $\mu$Jy $\rm beam^{-1}$) by approximately 2$\sigma$. The peak positions suggest an offset of 0.8$\arcsec$ ($\sim$17 au) from the star position. These features imply a potential asymmetry or eccentricity within the disc.

\begin{figure*}
	\includegraphics[width=1.0\textwidth]{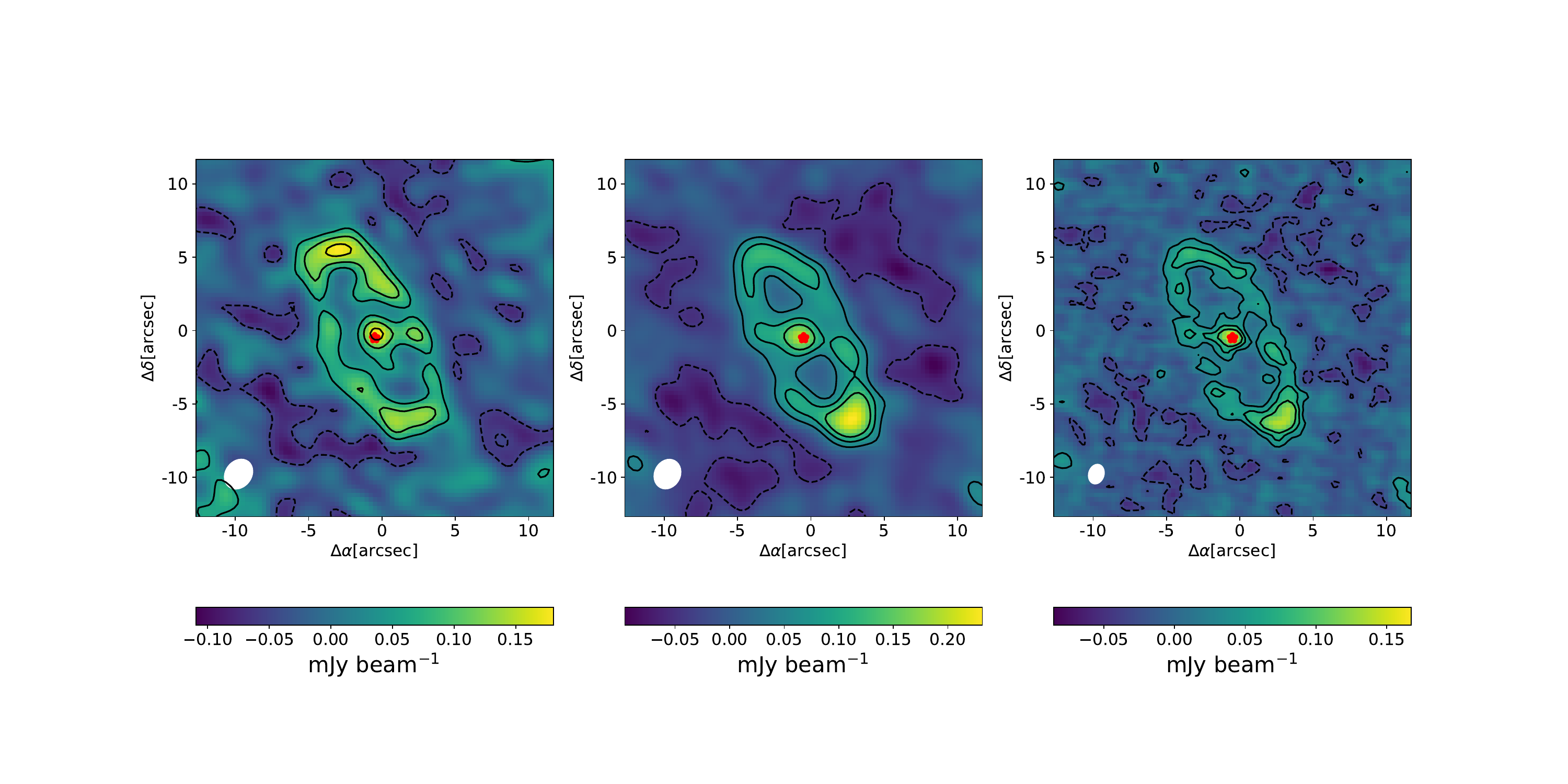}
	\caption{Continuum images of the two observation blocks with natural weighting. The left panel is for OB1 on 2017 March 28, and the right panel is for OB2 on 2018 May 9. The middle panel represents the UV tapering ($1.8\arcsec \times 1.5\arcsec$) image of OB2 for matching the beam size of OB1. Contours are drawn at the levels of $[-2,2,4,6]\times\sigma_{rms}$(left: $\sigma_{rms}=27\mu$Jy $\rm beam^{-1}$, middle: $\sigma_{rms}=22\mu$Jy $\rm beam^{-1}$, right: $\sigma_{rms}=15\mu$Jy $\rm beam^{-1}$) to better highlight the disc structure. The beam size is represented by a white ellipse (left: $2.24\arcsec \times 1.84\arcsec$, middle: $2.28\arcsec \times 1.89\arcsec$, right: $1.43\arcsec \times 1.12\arcsec$) in the bottom-left corner. The star location is marked with a red star symbol.}
	\label{fig:individual_c}
\end{figure*}

Figure~\ref{fig:individual_c} presents individual images of each observation block. To assess any potential asymmetry and verify the authenticity of the flux distribution, we conducted a comparative analysis of these images. We first created a version of the OB2 image with the same beam size as OB1 by smoothing the OB2 image with UV tapering. The smoothed OB2 image is shown in the middle panel of the figure. We observed from the left panel and middle panel that both datasets exhibit radiation originating from the central star and a debris disc, displaying similar morphologies. 
However, in the OB1 image, the peak of the central emission appears to be slightly northwest relative to the phase centre by 0.3$\arcsec$. Furthermore, there are differences in the brightness distribution of the debris disc between the two panels. In the OB1 image in the left panel, the disc flux distribution appears relatively uniform, with the brightest peak located in the northeast direction ($182 \pm 27$ $\mu$Jy $\rm beam^{-1}$)---higher than the opposite {\it ansa} ($143 \pm 27$ $\mu$Jy $\rm beam^{-1}$) of 1.5$\sigma$. In contrast, the middle panel of the UV tapering image from OB2 reveals significantly higher brightness levels in the southwestern region ($232 \pm 22$ $\mu$Jy $\rm beam^{-1}$) of -4.7$\sigma$ compared to the opposite {\it ansa} ($128 \pm 22$ $\mu$Jy $\rm beam^{-1}$). In the right panel, the unsmoothed OB2 image also shows a 3$\sigma$ difference between the two {\it ansa} $89 \pm 15$ $\mu$Jy $\rm beam^{-1}$ and $140 \pm 15$ $\mu$Jy $\rm beam^{-1}$, respectively. We also find the pixel with peak flux near the stellar position and report this flux as a measure of the central emission in the three images shown in Fig.~\ref{fig:individual_c}. The estimated central flux values are $181 \pm 27$ $\mu$Jy $\rm beam^{-1}$, $197 \pm 22$ $\mu$Jy $\rm beam^{-1}$, and $167 \pm 15$ $\mu$Jy $\rm beam^{-1}$, respectively.

\section{Modelling}
\label{sec:Model}

Based on the observations presented in the previous section, we conduct model analyses in the following subsections. In Sect.~\ref{sec:SED}, we utilise the SED to determine whether the disc comprises multiple components. In Sect.~\ref{sec:fit}, we employ a parametric model to fit the ALMA visibilities and constrain the dust density distribution within the debris disc. In Sect.~\ref{sec:var}, we separately fit the images from the two observation blocks to consider the possibility of any variability and assess the significance of asymmetries.

\subsection{Spectral energy distribution}
\label{sec:SED}

\begin{table}[htbp]
	\centering
	\caption{Photometry of HD 105211.}
	\label{tab:flux}
		\begin{tabular}{lcccc}
		\hline
		Wavelength & Flux & Used in Modelling & Reference \\
		($\mu m$) & (Jy) & & \\
		\hline
		0.349 & $24.02\pm0.24$ & Yes & 1 \\
		0.411 & $62.40\pm0.62$ & No &2 \\
		0.42 & $60.40\pm0.84$ & Yes & 3 \\
		0.51 & $68.25\pm0.64$ & Yes & 5 \\
		0.53 & $79.86\pm0.79$ & Yes & 3 \\
		0.54 & $79.82\pm0.45$ & Yes & 6 \\
		0.55 & $81.23\pm1.49$ & Yes & 4 \\
		0.64 & $82.15\pm0.79$ & Yes & 5 \\
		0.78 & $79.52\pm0.76$ & Yes & 5 \\
		1.2 & $64.72\pm15.69$ & Yes & 7 \\
		1.6 & $53.69\pm11.44$ & Yes & 7 \\
		2.2 & $34.84\pm7.55$ & Yes & 7 \\
		3.4 & $15.45\pm1.45$ & No & 8 \\
		4.6 & $7.24\pm0.31$ & No & 8 \\ 
		9 & $2.69\pm0.05$ & Yes & 9 \\
		12 & $1.46\pm0.02$ & Yes & 8 \\
		18 & $0.64\pm0.02$ & Yes & 9 \\
		22 & $0.42\pm0.01$ & Yes & 8 \\
		24 & $0.38\pm0.01$ & Yes & 10 \\
		70 & $0.47\pm0.04$ & No & 10 \\
		70 & $0.73\pm0.06$ & Yes & 11 \\ 
		100 & $0.73\pm0.10$ & Yes & 11 \\
		160 & $0.57\pm0.37$ & Yes & 12 \\
		1290 & $2.5\pm0.1\times10^{-3}$ & Yes & 13 \\ 
		\hline
	\end{tabular}
	\tablebib{(1)~\cite{2006yCat.2168....0M}; (2) \cite{2015A&A...580A..23P}; (3)\cite{2000A&A...355L..27H}; (4) \cite{1990A&AS...83..357B};
		 (5) \cite{2018yCat..36160012G}; (6) \cite{1997yCat.1239....0E}; (7) \cite{2003tmc..book.....C}; (8) \cite{2010AJ....140.1868W}; 
		 (9) \cite{2010A&A...514A...1I}; (10) \cite{2014ApJS..211...25C}; (11) \cite{2017MNRAS.468.4725H}; (12) \cite{2017arXiv170505693M};(13) this work.
	}
\end{table}

To construct the SED of HD105211, we collected observational data from literature papers, including photometric measurements from optical to millimetre wavelengths (see Table~\ref{tab:flux}). We also included photometry of the unresolved component and a \textit{Spitzer} InfraRed Spectrograph \citep[][IRS]{2004ApJS..154...18H} low-resolution spectrum spanning 7.5 to 38 $\mu$m, which was obtained from the CASSIS archive \citep{2011ApJS..196....8L}. We note that the ALMA Band 6 flux was obtained from our best-fit parameters including the disc flux ($2.34 mJy$) and flux at the stellar position (183 $\mu$Jy; see Table~\ref{tab:variability}). After that, the stellar and disc models were interpolated to the wavelengths of all data photometric points, including IRS wavelengths, and then minimisation was done across all photometric and spectral points. The stellar photospheric radiation was fitted using the PHOENIX model grid \citep{2013A&A...553A...6H}, while a blackbody radiation model was employed to account for the disc emission. 

We employed the method described in \citet{2019MNRAS.488.3588Y} to model the dust as a modified blackbody and fit it using the \texttt{MULTINEST} Markov Chain Monte Carlo (MCMC) algorithm \citep{2009MNRAS.398.1601F}. The emission spectrum includes wavelengths much longer than the typical sizes of particles contributing to the emission. Therefore, the blackbody function was multiplied by a factor $(\lambda_{0}/\lambda)^{\beta}$ for wavelengths $\lambda \geq \lambda_{0}$. Indeed, most debris discs display an emission spectrum steeper than the Rayleigh-Jeans tail at long wavelengths. We found that a single temperature model cannot fit the shape of the mid-IR emission in our SED, so an additional warm component was needed to fit the IRS spectrum. Hence, we adopted a two-temperature model to characterise the infrared excesses.

\begin{figure}
	\includegraphics[width=0.48\textwidth]{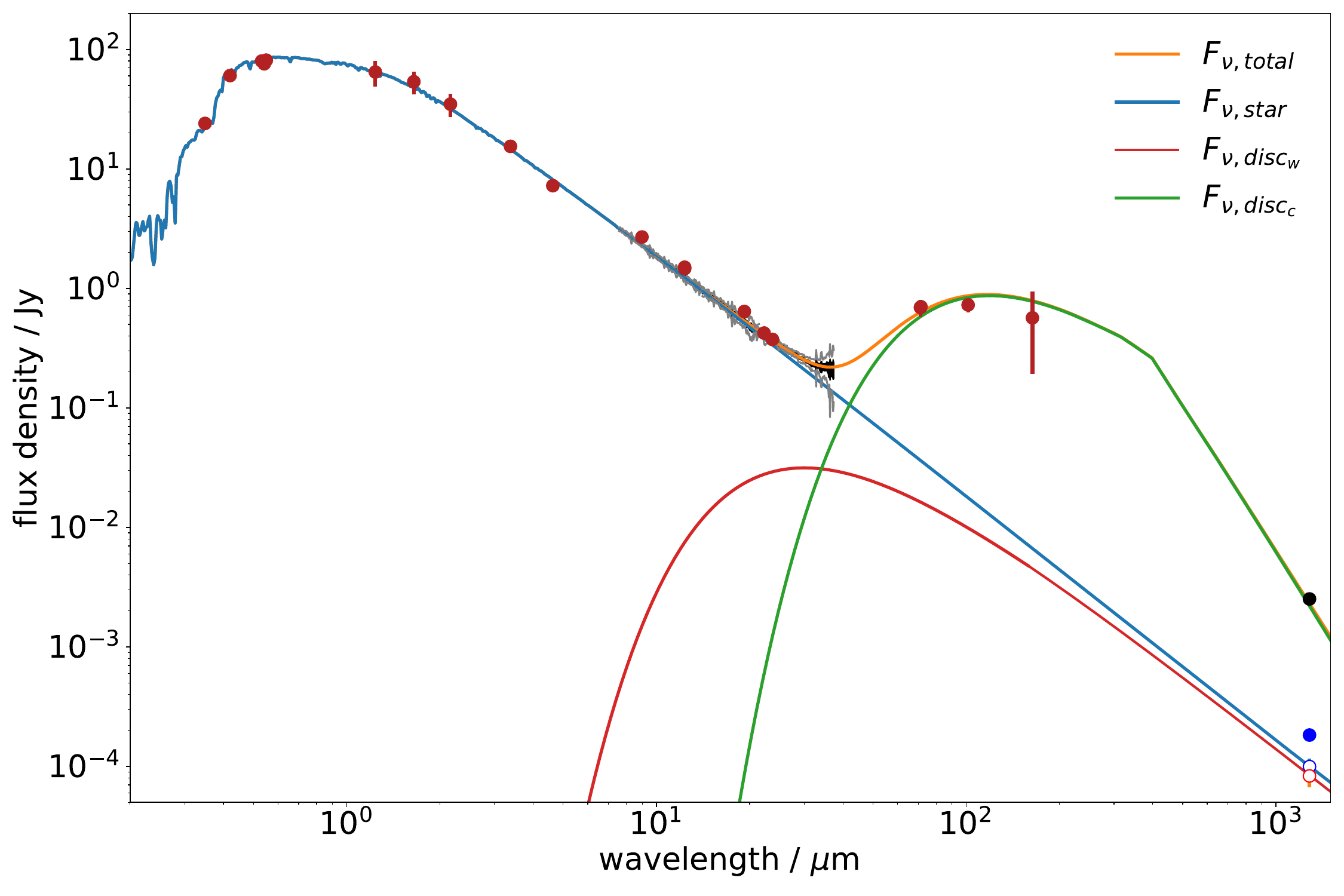}
	\caption{Spectral energy distribution of HD 105211. The photometry from the literature is plotted as red circles. The new ALMA observation is marked as a black circle, and the blue circle represents the flux at the stellar position. 
	The blue open circle donates the expected stellar flux, and the red open circle donates the flux seen at the stellar position after subtracting the predicted stellar flux. The shaded black line corresponds to the \textit{Spitzer} IRS spectrum, and the grey line represents its $\pm3\sigma$ uncertainties. The blue line represents the stellar photosphere emission. The green and red lines denote disc emission of the cold and warm components, respectively, whilst the orange line denotes their total emission.}
	\label{fig:sed}
\end{figure}

Our SED results are presented in Fig.~\ref{fig:sed}. The model fit yielded the effective temperature of the star ($T_{eff}$) as $6990\pm100$ K and stellar luminosity ($L_{\star}$) as $6.8\pm0.05$ $L_{\sun}$. For the debris disc components, we derived $T_{bb}$ = $43\pm1$ K, $R_{bb}$ = $110\pm5$ au and $L_{disc}$/$L_{\star}$ = $6.3\pm0.3\times10^{-5}$ for the cold disc and $T_{bb}$ = $170\pm20$ K, $R_{bb}$ = $7\pm1$ au and $L_{disc}$/$L_{\star}$ = $9\pm1\times10^{-6}$ for the warm disc. Based on the best-fit SED model, the estimated stellar flux at ALMA 1.3 mm is $100\pm2.5 \mu$Jy. The modified blackbody parameters for the cold and warm components are $\lambda_{0} = 380\mu$m, $\beta = 2.3$, and $\lambda_{0} = 160\mu$m, $\beta\sim0$, respectively. In this fit the warm component is constrained to have the characteristics of a pure blackbody, as it is required to match the millimetre flux of the central component observed in the ALMA images (see Sect.~\ref{sec:results}).

\subsection{Image modelling}
\label{sec:fit}

\subsubsection{Parametric model}
\label{sec:Parametric model}

We fit a parametric disc model to place more constraints on the disc as an optically thin torus using observed visibilities, which is commonly used for describing millimetre-resolved debris discs. We averaged the data into 10-second long chunks and spectrally averaged the four spectral windows down to four channels per spectral window. Then we used the CASA "statwt" task to recompute the visibility weighting following the averaging process.

We used the parametric structure and modelling code described in \citet{2018MNRAS.475.4924K}, combining a simple geometrical model similar to "zodipic" (\citep{2012ascl.soft02002K}) and a MCMC fitting procedure in the visibility space. The image generated was then Fourier transformed to the visibility plane and sampled at the same UV points as our continuum observations, computing the $\chi^{2}$ between the model and the data to quantify the goodness-of-fit \citep{2018MNRAS.476.4527T}. We assumed posterior distribution functions are uniform for all parameters.

To find the best-fitting model and explore the parameter space for posterior probabilities, we used the ensemble MCMC method proposed by \cite{2010CAMCS...5...65G} and as implemented by the \texttt{EMCEE} package \citep{2013PASP..125..306F}. 
The model was initialised near the optimal solution based on prior testing runs. We used 100 walkers and chains with 1000 steps and discard the first 900 steps as a ‘burn-in’ phase to reach the stable stage. 

Our model is an axisymmetric Gaussian ring of radius $r_{0}$, for which the additional parameters are the radial $\sigma_{r}$ and the vertical $\sigma_{h} $ density dispersions. The vertical density $\sigma_{h}$ is the vertical aspect ratio (H/r) in a Gaussian distribution from the midplane. The model has other parameters, such as disc flux density \textit{F}, disc radius $r_{0}$, disc position angle \textit{PA} (east of north), disc inclination \textit{i}, and small sky offset of the disc from the phase centre $x_{0}$, $y_{0}$. In Sect.~\ref{sec:Observations}, we found that the continuum images present seemingly unresolved emission from close to the location at the central star, so we added another parameter: the flux density at stellar position $F_{c}$. The dust density is defined by Eq~(\ref{equation:1}):

\begin{equation}
	\rho_{dust}(r,z)=\rho_{0}exp[-\frac{(r-r_{0})^2}{2\sigma_{r}^2}-\frac{z^2}{2{r}^2\sigma_{h}^2}].
	\label{equation:1}
\end{equation}

Since the surface brightness profile indicates that the disc might be eccentric, we also considered an eccentric model with the same parameters as the Gaussian model to fit the image while adding an additional parameter eccentricity \textit{e}. 
For the eccentric model, there are two main effects: the stellar position that is offset from the disc centre, and the apocentre glow \citep{2016ApJ...832...81P}. Therefore, we replaced $\rho_{0}$ and $r_{0}$ in Eq.~(\ref{equation:1}) with formulas of \citet{2017MNRAS.465.2595M}: 

\begin{equation}
	\rho_{c}=\rho_{0}[1-e\cos(\theta)],
	\label{equation:2}
\end{equation}

\begin{equation}
	r_c=\frac{r_0(1-e)^2}{1+e\cos(\theta)},
	\label{equation:3}
\end{equation}

where \textit{e} represents the eccentricity of the disc and $\theta$ denotes the azimuthal angle of particles measured from the pericentre, which is not shown in the list of parameters.

\subsubsection{Results}
\label{sec:results}

The best-fitting parameters for both models are given in Table~\ref{tab:variability}
, and the posterior distributions for all parameters are presented in Appendix~\ref{sec:pos}. The Gaussian model yields a disc radius of $133.7 \pm 1.6$ au, with a width of $23.6 \pm 4.6$ au, an inclination of $67.0 \pm 0.7\degr$, and a position angle of $28.3 \pm 0.7\degr$. The eccentric model provides geometric parameters that are nearly identical to those obtained from the Gaussian ring model, yielding a disc radius of $133.3 \pm1.4$ au with a width of $24.1 \pm 5.1$ au and an inclination of $66.9 \pm 0.7\degr$.

Regarding the constraints on disc thickness, both models obtained the same vertical aspect ratio of 0.07. Considering the shape of the disc and the size of the beam, the vertical direction may not be resolved. However, our model-fitting results still align closely with the typically assumed value of 0.05 for the vertical aspect ratio of the debris disc. 
Additionally, the stellar fluxes of the two models are $183\pm 14$ $\mu$Jy and $183\pm 15$ $\mu$Jy, which are consistent with the measured flux $168\pm 16$ $\mu$Jy within the margin of error but higher than the stellar flux by about 5 $\sigma$. 

Based on our investigation, as discussed in Sect.~\ref{sec:ob Results}, we inferred a possible eccentricity of the disc from the deviation of peak flux and peak radius in the radial profile. However, the eccentricity obtained from the eccentric model was found to be $e\leq0.02$. Consequently, the geometric shape of the disc in both models was nearly identical, suggesting that the disc is an axisymmetric Gaussian narrow ring. Any asymmetry seen in the disc images cannot be attributed to the eccentricity.

\begin{figure}
	\centering
	\includegraphics[width=0.45\textwidth]{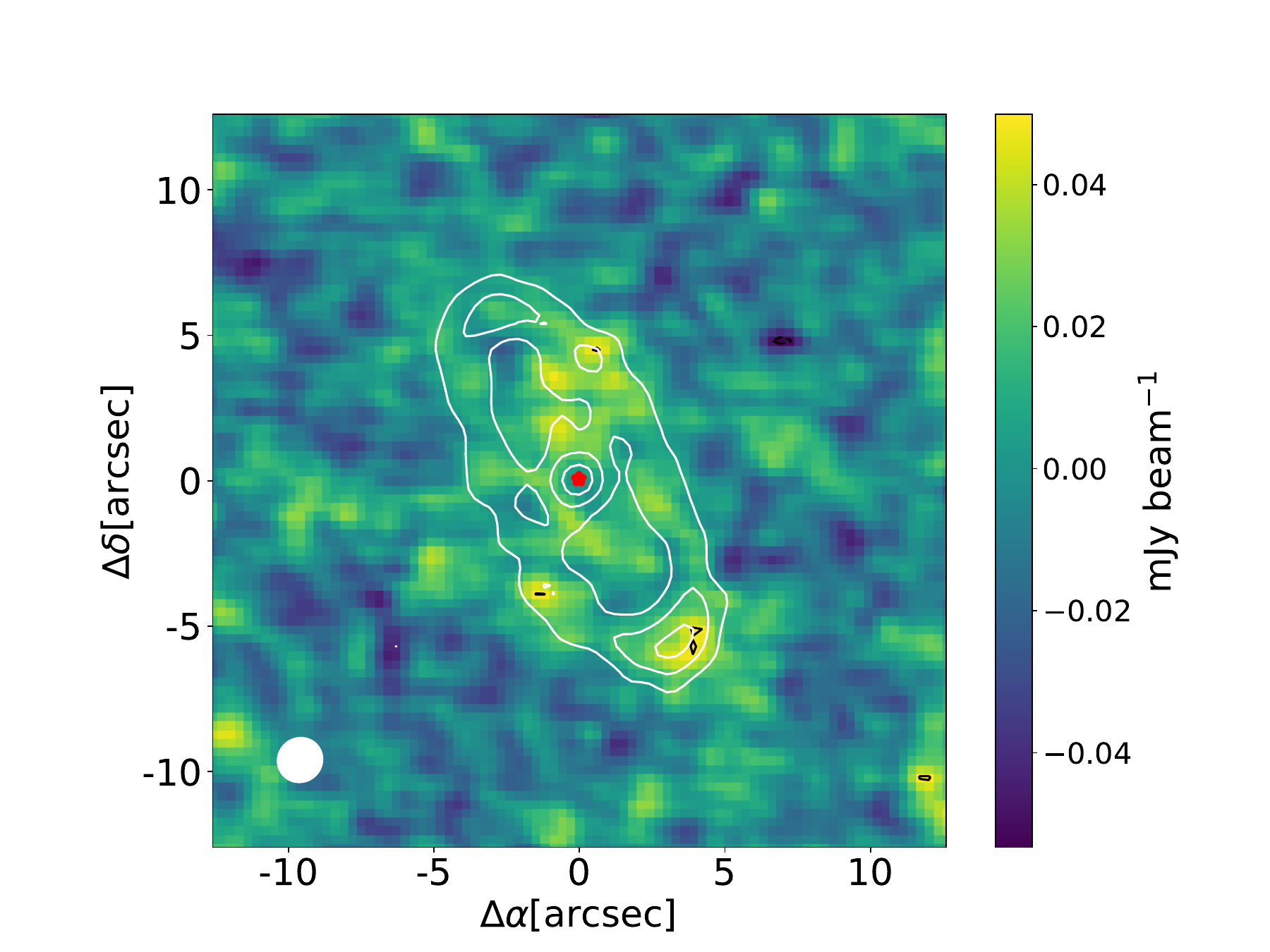}
	\caption{Residual of the Gaussian model using the same contour levels and beam size as the continuum image with $3 \times \sigma_{rms}$($\sigma_{rms}$= 16$\mu$Jy $\rm beam^{-1}$) and beam size in $1.62\arcsec \times 1.57\arcsec$. The red star symbol is the star's location. We added contours of the continuum image (white) for comparison with the residual image.}
	\label{fig:residual}
\end{figure}

Since the two models had almost the same geometric shape, we only computed the residual maps of the Gaussian model at the same observational resolution, which is shown in Fig.~\ref{fig:residual}. The residual maps exhibit three significant (3$\sigma$) residual clumps coincident with the ring, indicating the higher dust density or noise level for these regions. There are also 2$\sigma$ residual clumps at the nor-theast of the star that need further confirmation. And the residual around the star is lower than 2$\sigma$, suggesting the extension in brightness profile may be attributed to the noise.

\subsection{Variability}
\label{sec:var}

In this section, we perform parameter model fitting on the data of the two observations (OB1 and OB2) separately to quantify the impacts of noise and variability on the results. We used both the Gaussian and eccentric models and show the results in Table~\ref{tab:variability}.

\begin{table*}
	\centering
	\caption{Best-fitting parameters of each observation.}
	\label{tab:variability}
	\resizebox{180mm}{!}{
		\begin{tabular}{lccccccc} 
			\hline
			 & & \multicolumn{2}{c}{Combination} & \multicolumn{2}{c}{2017 Mar. 28}  & \multicolumn{2}{c}{2018 May 9}  \\
			\cline{3-8}
			Parameter & Description & Gaussian & Eccentric & Gaussian & Eccentric & Gaussian & Eccentric \\
			\hline
			$x_{0}$(\arcsec) & R.A. offset & $0.021\pm 0.032$ & $0.030\pm 0.034$ & $0.049\pm 0.062$ & $0.059\pm 0.070$&  $0.009\pm 0.038$ &  $0.036\pm 0.041$\\
			$y_{0}$(\arcsec) & Dec. offset & $0.052\pm 0.031$ & $0.061\pm 0.035$ & $0.153\pm 0.082$ & $0.184\pm 0.099$ &  $0.006\pm 0.040$ &  $0.054\pm 0.042$\\
			$r_{0}$(au) & disc central radius & $133.7\pm 1.6$ & $133.3\pm 1.4$ & $134.3\pm 2.6$ & $134.5\pm 2.6$ &  $132.9\pm 1.6$ &  $132.5\pm 1.6$\\
			$\sigma_{r}$(au) & Radial standard deviation & $10.0\pm 2.1$ & $10.2\pm 2.2$ & $9.3\pm 4.3$ & $9.3\pm 4.7$ &  $10.0\pm 2.0$ &  $10.0\pm 2.0$\\
			$\sigma_{h}$ & Vertical aspect ratio & $0.07\pm 0.02$ & $0.07\pm 0.01$ & $0.06\pm 0.03$ & $0.06\pm 0.03$ &  $0.08\pm 0.02$ & $0.08\pm 0.02$\\
			$F(mJy)$ & Flux of disc & $2.34\pm 0.13$ & $2.32\pm 0.13$ & $2.35\pm 0.21$ & $2.36\pm 0.21$ &  $2.40\pm 0.16$ & $2.41\pm 0.16$\\
			$F_{c}(\mu Jy)$ & Flux at stellar position& $183\pm 14$ & $183\pm 15$ & $198\pm 30$ & $200\pm 31$ &  $184\pm 16$ & $184\pm 16$\\
			PA(\degr) & disc position angle & $28.3\pm 0.7$ & $28.3\pm 0.7$ & $27.9\pm 1.0$ & $28.0\pm 1.1$ &  $28.5\pm 0.8$ & $ 28.2\pm0.8$\\
			i(\degr) & disc inclination &  $67.0\pm 0.7$ & $66.9\pm 0.7$ & $65.8\pm 1.1$ & $65.9\pm 1.1$ &  $67.9\pm 0.8$ & $67.8\pm 0.8$\\
			e & disc eccentricity & - & $0.014\pm 0.007$ & - & $0.004\pm 0.010$ & - & $0.028\pm 0.011$\\
			\hline
		\end{tabular}
	}
\end{table*}

Although there is an asymmetry in the brightness distribution of the continuum image for both OB1 and OB2, the eccentricity obtained from the eccentric model fitting is lower than 0.03 for both observations. Both the eccentric models and Gaussian models' geometric shapes closely align, indicating that these brightness asymmetries may not originate from eccentric structures. Therefore, we have chosen the Gaussian model parameters as our result. These parameters of both observations are consistent with the results obtained in Sect.~\ref{sec:fit}, allowing us to better model the shape of the HD105211 debris disc. We note that the stellar fluxes obtained from the fitting of OB1 and OB2 are $198\pm 30$ $\mu$Jy and $184\pm 16$ $\mu$Jy, respectively, matching the measured values in the OB1 and OB2 image of $181 \pm 27$ $\mu$Jy and $167 \pm 15$ $\mu$Jy within the margin of error. These flux values are significantly higher than the expected flux of 100 $\mu$Jy based on SED by about 4 $\sigma$ and 5 $\sigma$ respectively, confirming that the excess flux obtained in Sect.~\ref{sec:fit} is not due to noise or stellar variability.

From the continuum image of OB1, we found that the star deviated from the phase centre by 0.3$\arcsec$. The best-fit offset for the $x$ and $y$ directions are $0.05\pm0.06\arcsec$ and $0.15\pm0.08\arcsec$, respectively. By checking the phase centres of the two observation blocks, we found that the phase centres of the two observation blocks deviated by 0.02$\arcsec$ in the declination direction. While the star is moving in the southeast direction at a rate of 0.034 arcsec/yr \citep{2007A&A...474..653V}, consistent with the direction of deviation in the observations, this would have resulted in an offset that is smaller than the displacement obtained from the fitting. This displacement is likely caused by the absolute pointing precision of ALMA of around 10\% of the beam since a ~1.5$\arcsec$ beam makes an offset of 0.15$\arcsec$ not very significant.

\begin{figure*}
	\includegraphics[width=1.0\textwidth]{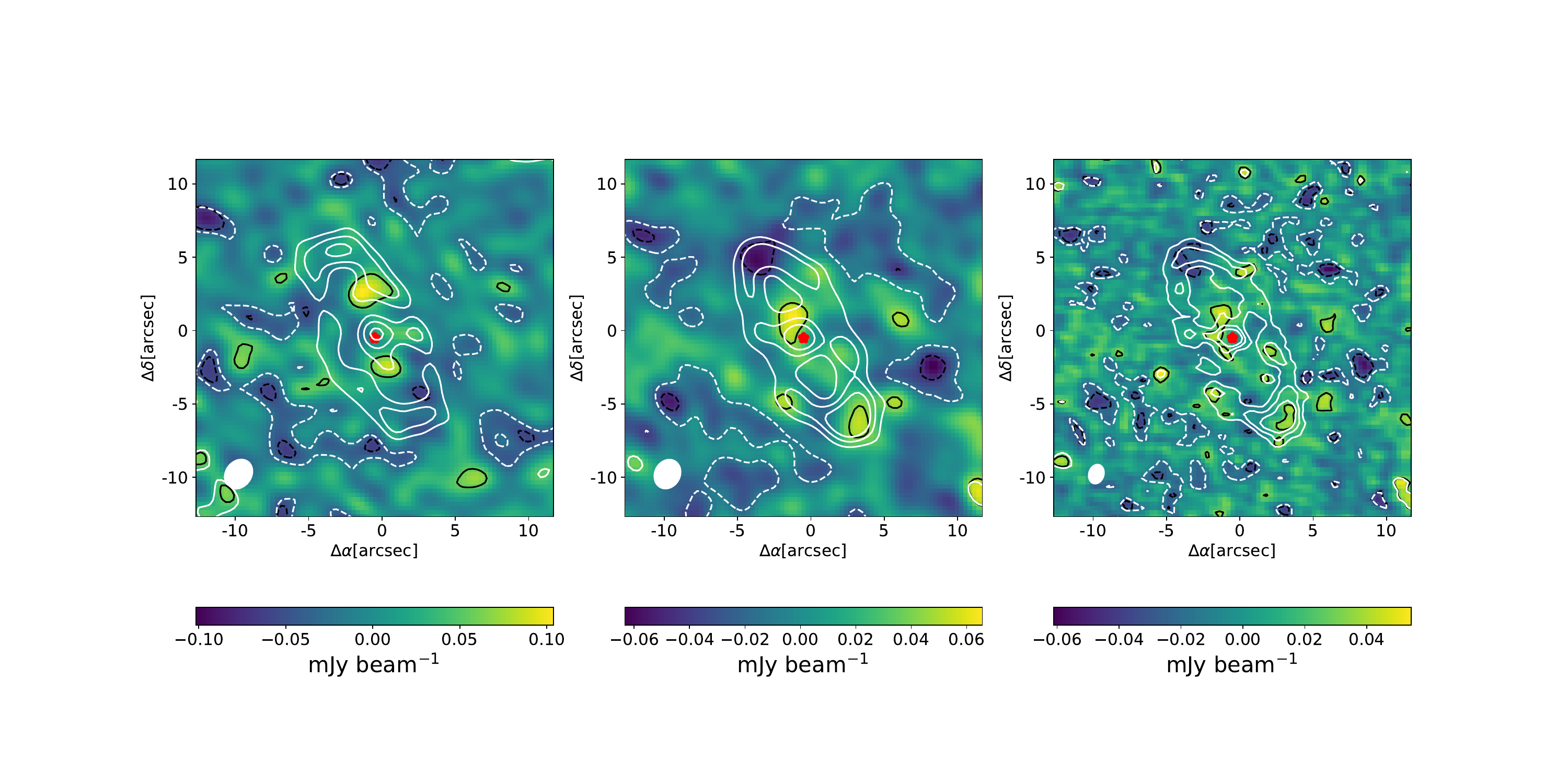}
	\caption{Residual maps using the same contour levels ($[-2,2,4,6]\times\sigma_{rms}$, left: $\sigma_{rms}=27\mu$Jy $\rm beam^{-1}$, middle: $\sigma_{rms}=22\mu$Jy $\rm beam^{-1}$, right: $\sigma_{rms}=15\mu$Jy $\rm beam^{-1}$) and beam sizes (left: $2.24\arcsec \times 1.84\arcsec$, middle: $2.28\arcsec \times 1.89\arcsec$, right: $1.43\arcsec \times 1.12\arcsec$) as Fig.~\ref{fig:individual_c}. We added contours of the continuum image (white) to compare the residual positions.}
	\label{fig:individual}
\end{figure*}

Figure~\ref{fig:individual} presents the residuals from the individual Gaussian model results of OB1 (left) and OB2 (right) and UV tapering of OB2 (middle). We have also overlaid the contours of the continuum images for comparison with the residual positions. In all panels, the residuals are consistent with the residual of the combined image, showing two clumps on the ring and another clump in the northeast direction near the star, matching the contours in the individual continuum images. Although there are still some differences in the image residuals of different observation blocks, most of the lower-level residuals can be attributed to noise. However, the significant difference between the negative residual in the NE (-2$\sigma$) and the positive residual in the SW (2$\sigma$) of both OB2 and the UV tapering suggests significant brightness asymmetry distribution in OB2. There is also the presence of more than one clump on the minor axis of the disc near the star in OB2. Considering the projection of the disc, these residuals could be the noise or clump rather than the extension of the inner component.

\section{Discussion}
\label{sec:discussion}

\subsection{Comparison with \textit{Herschel}}

To further refine the characterisation of the HD105211 debris disc, we compared the structure and geometry of our ALMA best-fit model with previous imaging studies at shorter wavelengths. \citet{2017MNRAS.468.4725H} deconvolved the star and the disc emission to obtain resolved images in the \textit{Herschel}/PACS 70, 100, and 160 $\mu$m bands. These images revealed extended emission from a disc with a width of 100 $\pm$ 20 au starting at an inner edge of 87 $\pm$ 2.5 au. The disc had an inclination of 71 $\pm$ 2\degree and a position angle of 30 $\pm$ 1\degree. In contrast, our ALMA best-fit model for the disc is a narrow ring with a width of 24.6 $\pm$ 4.6 au located at 133.7 $\pm$ 1.6 au and with an inclination of 67 $\pm$ 0.7\degree, and its position angle is 28.3 $\pm$ 0.7\degree.

\begin{figure}
	\centering
	\includegraphics[width=0.5\textwidth]{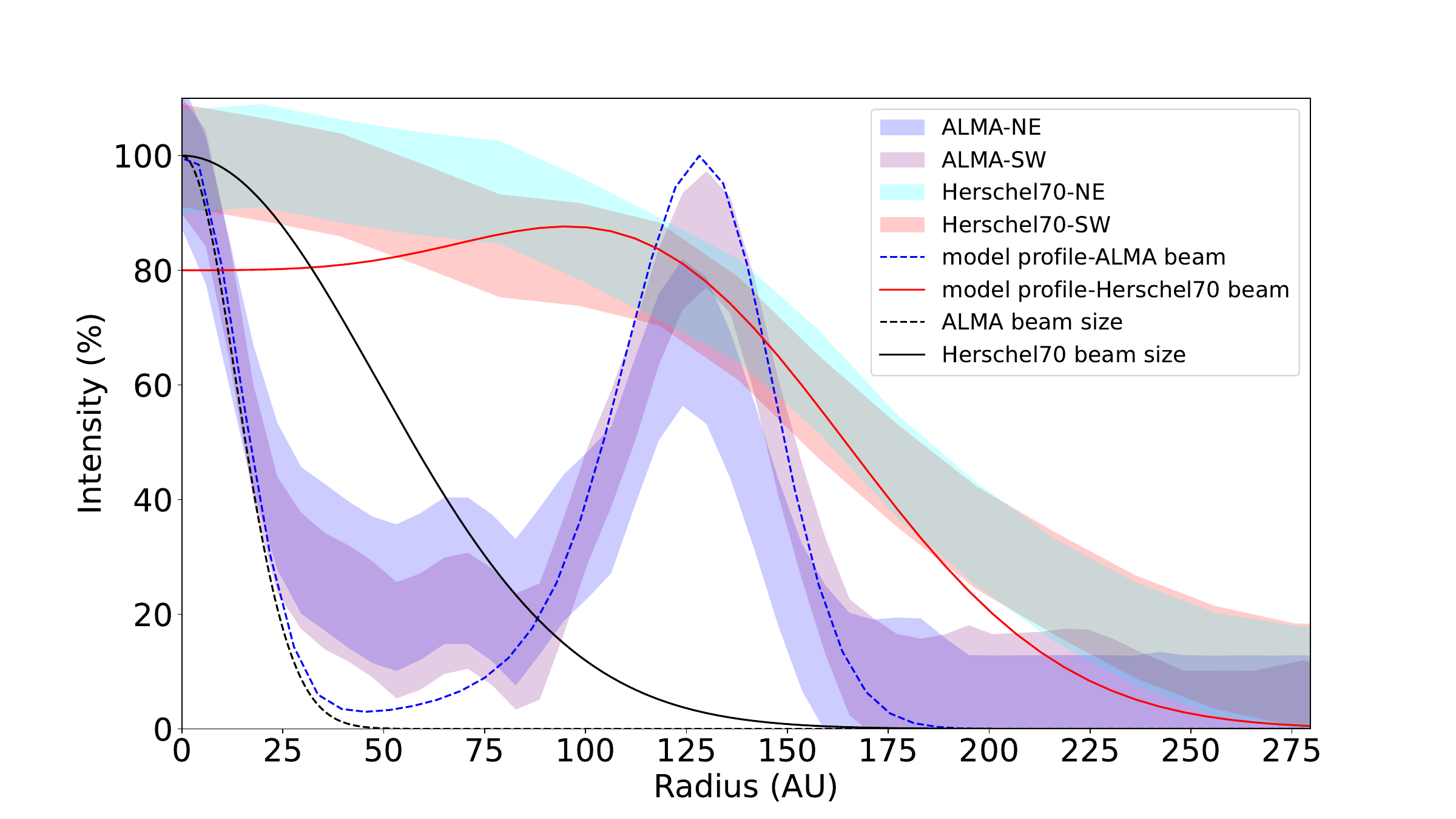}
	\caption{Surface brightness profile of the disc along the major axes in the ALMA and original \textit{Herschel} 70 $\mu$m images. The shaded regions correspond to the 1 $\sigma$ uncertainties. The blue dashed line represents the radial profile of our best-fit parametric model convolved with the ALMA synthesised beam, while the red line represents the radial profile of the model convolved with the \textit{Herschel} 70 $\mu$m beam. The black line corresponds to the \textit{Herschel} 70 $\mu$m beam size, and the black dashed line denotes the ALMA beam size.}
	\label{fig:profile_c}
\end{figure}

We also compared the radial profiles between ALMA and the \textit{Herschel} 70 $\mu$m original image in Fig.~\ref{fig:profile_c}. The blue dashed line represents the radial profile of our best-fit parametric model convolved with the ALMA synthesised beam, while the red line represents the radial profile of the model convolved with the \textit{Herschel} 70 $\mu$m beam. The radial profile of the \textit{Herschel} 70 $\mu$m observed image in the figure shows emissions higher near the stellar position than the model, which may arise from the warm component not included in the model and the stellar emission varying at different wavelengths. However, emissions near the inner edge of the ring are still significantly high, indicating a more extended inner edge of the disc in the \textit{Herschel} image. Similarly, the observed profile exhibits more extended emissions at the outer edge, suggesting a broader outer edge of the ring in the \textit{Herschel} observation. Their model, combined with SED, yielded a disc width of 100 au, which was also significantly wider than the results from ALMA observations.

The \textit{Herschel} observations presented a closer inner edge and a further outer edge, suggesting that there might be a different radial distribution of dust between the infrared and millimetre wavelengths for the reason that they trace grains of different sizes \citep[see. Fig 5 of ][]{2002MNRAS.334..589W}. Larger grains are primarily influenced by gravity, making millimetre observations suitable for tracing the planetesimal belt where dust is generated. In contrast, smaller grains are affected by non-gravitational forces, especially stellar radiation pressure and P-R drag. These forces can extend the dust distribution from the planetesimal belt, creating an extended structure visible at shorter wavelengths. Consequently, the broad structure of the Herschel observations sees smaller grains whose distribution is modified by radiation forces, and the new ALMA results provide a more accurate description of the underlying planetesimal belt structure.

\citet{2017MNRAS.468.4725H} also showed an asymmetry between the two sides of the disc after deconvolution, interpreted as ansae lying at different radii, from which they inferred a disc eccentricity of 0.20 $\pm$ 0.03. However, their analysis also led to the conclusion that the structure and emission shown in the \textit{Herschel} images could be effectively modelled with an axisymmetric annulus. There is also asymmetry in the ALMA images, as described in Fig.~\ref{fig:profile}, which illustrated the offset of brightness peaks in OB1 and variations in intensity in OB2. The offset of brightness peaks may be attributed to the offset of the star to the phase centre, and the significant (4$\sigma$) variations in intensity may be caused by a transient collision event or noise that needs further observations. Our best-fitting model indicates a disc eccentricity below 0.03, suggesting that the disc does not exhibit asymmetry caused by eccentricity.

\subsection{Collisional status of the disc}
\label{sec:State}

The new high-resolution ALMA map reveals the structure of the debris disc around HD105211, and  it allowed us to investigate the detailed characteristics of the disc with these geometrical parameters. As larger bodies in the debris disc are destroyed, they give rise to the formation of the smallest dust grains. These small particles are then blown out from the disc by radiation pressure, leading to a gradual loss of mass over time. As a result, older debris discs exhibit reduced luminosities and become more challenging to be detected compared to younger discs. By assuming a steady-state collisional cascade model, we could estimate the mass-loss rate of the smallest grains using a simple equation provided by \citet{2017ApJ...842....9M} and given as $\dot{M}_{D_{min}}=1.2\times10^{3}R^{1.5}\bigtriangleup R^{-1}f^{2}L_{\star}M_{\star}^{-0.5}$, where \textit{R} is in astronomical units, \textit{L} is in $L_{\sun}$, and \textit{M} is in $M_{\sun}$. 

Using the best-fitting parameters and fractional luminosity obtained in Sect.~\ref{sec:Model}, we could derive a minimum grain size mass-loss rate of $\dot{M}_{D_{min}}=1.7\times10^{-3}$ $M_{\oplus}$ $Myr^{-1}$ for the disc. 
This relatively low mass-loss rate of small dust particles implies a commensurately low mass of such particles within the disc. Despite the considerable age of HD105211, estimated to be from 1.34 Gyr to 1.92 Gyr, the cumulative loss of mass remains limited, ranging from 2.2 $M_{\oplus}$ to 3.2 $M_{\oplus}$. However, it is crucial to acknowledge that the mass-loss rate is not a static constant throughout the evolution of the debris disc. The equation $\dot{M}_{D_{min}} \propto f^{2}$ within the framework emphasises the pronounced impact of fractional luminosity on the mass-loss rate, with the low fractional luminosity of HD105211 emerging as the primary contributing factor to the inferred low mass-loss rate.

Assuming a millimetre dust opacity of $\kappa=2.3$ $cm^{2}$ $g^{-1}$, the mass of millimetre-sized dust can be calculated using equation (5) of \citet{2008ARA&A..46..339W} and the best-fitting flux. The calculated mass of millimetre-sized dust in the debris disc is $1.1\times10^{-2}$ $M_{\oplus}$. The dust mass obtained in the model results by \citet{2017MNRAS.468.4725H} is $2.45\times10^{-2} M_{\oplus}$, slightly greater than our results. Their model combining the SED may be affected by warm components, such as the need for a greater mass and broader spatial distribution of dust in the model when dust is distributed over a more distant region. However, the estimation of dust mass is strongly dependent on the assumption about the opacity, which is influenced by differences in properties such as dust composition. Varying assumptions can also result in differences in estimating the dust mass.

Moreover, we determined the actual radius of the debris disc around HD105211, obtained from Sect.~\ref{sec:SED} and \ref{sec:fit}, to be $133.7\pm1.6$ au, while we calculated the blackbody radius to be $110\pm 5$ au. We found the ratio of the actual radius to the blackbody radius, denoted as $\Gamma$, to be $1.2\pm0.2$, significantly lower than the expected value of $2.8\pm0.1$ according to the model proposed by \citep{2015MNRAS.454.3207P}. This discrepancy suggests that the emission spectrum in the HD105211 debris disc is primarily governed by large particles, indicating a relatively lower abundance of smaller particles. For a narrow ring-like planetesimal belt, small grains are destroyed much more efficiently than they are created for a low excitation, leading to a depletion of this population \citep{2008A&A...481..713T}, which could be confirmed in the scattered light observations.
Remarkably, no observations of scattered light have been conducted for HD105211 thus far, which limits our ability to determine the level of small dust particles through scattered light imaging. This aspect warrants further investigation in future studies.

\subsection{Warm inner dust component}
\label{sec:Inner component}

All the ALMA images including the individual images of OB1, OB2, and the combined image exhibit excess emissions at the stellar location higher than expected from the star at greater than $3\sigma$. This is consistent with the emission expected from the warm component, at least if that emission falls off like a blackbody (and not steeper than that). For the warm components observed for the first time in ALMA imaging, we placed some constraints on their origin and properties based on the results obtained previously.

Drawing from the findings of \citet{2013MNRAS.433.2334K}, we considered two potential processes for this dust component. The first process involves an in situ planetesimal belt that is analogous to the solar system's asteroid belt. Within this belt, planetesimals may have remained in their original positions during the disc evolution \citep{2007ApJ...658..569W, 2008ARA&A..46..339W}. These planetesimals could have then undergone collisional cascades, ultimately producing the observed excess dust emission. The second process is comet delivery. In this case, planetesimals from more distant regions, analogous to the Kuiper Belt of the solar system, could have been transported to the inner region through interactions with planets \citep{2012A&A...548A.104B, 2012MNRAS.420.2990B}. This material could have then replenished the dust component, resulting in the observed excess radiation. 

Regarding in situ evolution, the disc would resemble a fairly normal warm asteroid belt analogue. We can compare the maximum fractional luminosity of steady-state collisional evolution at the blackbody radius to constrain the disc. Using equation (20) in \citet{2007ApJ...658..569W} and assuming the warm component blackbody radius $R_{bb}$ = $7\pm1$ au (see Sect.~\ref{sec:SED}) yields the maximum fractional luminosity of $(2.4\pm1.3)\times10^{-6}$. This is lower than the observed fractional luminosity of $(9\pm1)\times10^{-6}$, which suggests the belt should be farther out, closer to $13\pm2$au to have retained such a high fractional luminosity to the current age. This is consistent with the true radius of the disc being larger than its blackbody radius because small dust is always hotter than a blackbody\citep{2015MNRAS.454.3207P}. We note, however, that this introduces a tension with the blackbody spectrum inferred from the SED since to fit the dust temperature at this greater distance suggests the presence of small grains, and these would emit inefficiently at longer wavelengths. This could indicate that the disc is broad, but this cannot yet be concluded given the current constraints.
For the narrow belt to be at this radius, it would either require the particles to be very small, which is incompatible with a pure blackbody spectrum, or for the belt to be very broad, with the observed millimetre radiation coming from its outer edges.

Considering an alternative scenario, for a quasi-steady-state collisional cascade at a single temperature, \citet{2012ApJ...754...74G} showed a size distribution index of 3.65 and a spectral index of 2.65 in the Rayleigh-Jeans tail of the SED. 
This is inconsistent with our SED fitting results, where the warm component emission is well fitted by a nearly pure blackbody. In contrast, some debris discs do seem to have pure blackbody emission, such as $CP-72~2713$, AU Mic \citep{2020AJ....159..288M, 2015ApJ...811..100M}. 
As \cite{2015ApJ...811..100M} concluded, the disc of AU Mic may have a grain size distribution inconsistent with that expected from a quasi-steady-state collisional cascade. Since comets are the dominant source of the zodiacal cloud in our Solar System \citep{2022MNRAS.510..834R, 2010ApJ...713..816N}, comet fragmentation may also be the source of the inner clumps we observed. The radially extended nature perhaps can explain why the warm component radiation appears to be a pure blackbody spectrum, suggesting that comet-like delivery is a possible source of dust. This dust from comet delivery may provide a grain size distribution different from the steady-state collisional cascade. 

\section{Conclusions}
\label{sec:Conclusions}

In this paper, we have presented high-resolution observations of the HD 105211 debris disc at ALMA 1.3 mm for the first time. The resulting images offer clear insights into both the central star and the debris disc. Our analysis determined the disc inclination to be $67.0 \pm0.7\degree$, with a position angle of $28.3 \pm0.7\degree$, consistent with previous observations. Through the application of the fitting procedures to model the radial and vertical structure of the disc, we derived a narrow ring with a centre-to-star distance of $133.7\pm1.6$ au and a width of $23.6\pm4.6$ au. The ratio of the true disc radius to the blackbody radius is $1.2\pm0.2$, lower than the expectation of $2.8\pm0.1$ for a disc around a star of this luminosity, indicating that the disc is primarily composed of larger dust grains, with a limited presence of smaller particles.

The structure distribution in the ALMA images displays a narrow ring, different to the broad structure of Herschel images, which indicates the distribution of the smaller grains is modified by radiation forces. The brightness distribution in the ALMA images displays a potential asymmetry, with the various locations and intensity of brightness peak. However, our fitting of the eccentric disc model resulted in an eccentricity $e\leq0.03$, suggesting that the debris disc does not exhibit significant eccentricity. The different brightness distributions between the two observation blocks suggest the presence of potential transient collision events during the interval between these two observations or that the noise level in OB2 is particularly high at the ends of the projected major axis (-2$\sigma$ and 2$\sigma$, respectively).

By comparing our results to those from observations at different time points and examining the SED, we have detected warm dust surrounding the central star at a wavelength of 1.3 mm for the first time. We have also discussed the potential sources and characteristics of this warm dust component. The spectral index of the warm component is approximately two, indicating the spectrum is close to a pure blackbody. This differs from the expectation of a quasi-steady-state collisional cascade. Comparing the fractional luminosity of the warm component with the steady-state collisional cascade model shows that the warm component should be placed at 13 au. Considering the distribution of dust, we posit that this warm dust likely originates from both the in situ broad disc and comet delivery, hinting at the potential presence of planets formed through comet interactions. These possibilities could be tested with upcoming JWST observations (PI: Matrà Luca, Proposal ID: 5650). This discovery holds significant implications for our understanding of the origin and evolution of extrasolar warm dust and the formation of Earth-like planets.

\begin{acknowledgements}
The authors thank the referee for careful reading of the manuscript and feedback, which improved the clarity of this paper. We also like to thank Shane Hengst for providing the \textit{Herschel} original and deconvolved images, and the fruitful discussion on comparison between ALMA and \textit{Herschel} surface brightness profiles from Yinuo Han.

This work was supported in part by the NSFC under grants U1631109, 11703093, and U2031120. This work was also supported in part by the Special Natural Science Fund of Guizhou University (grant no. 201911A) and the First-class Physics Promotion Programme (2019) of Guizhou University. GMK is supported by the Royal Society as a Royal Society University Research Fellow.

This paper makes use of the following ALMA data: ADS/JAO.ALMA\#2016.1.00637.S. ALMA is a partnership of ESO (representing its member states), NSF (USA), and NINS (Japan), together with NRC (Canada), \textit{MOST} and ASIAA (Taiwan), and KASI (Republic of Korea), in cooperation with the Republic of Chile. The Joint ALMA Observatory is operated by ESO, AUI/NRAO, and NAOJ. The National Radio Astronomy Observatory is a facility of the National Science Foundation operated under a cooperative agreement by Associated Universities, Inc. 

This research use the \tt{PYTHON} packages \texttt{APLPY}, \texttt{ASTROPY} \citep{2012ascl.soft08017R}, \texttt{CORNER} \citep{corner}, \texttt{MATPLOTLIB} \citep{2007CSE.....9...90H}, \texttt{PANDAS} \citep{pandas}, \texttt{NUMPY} and \texttt{SCIPY} \citep{5725236}.

\end{acknowledgements}

\bibliographystyle{aa}
\bibliography{references}

\onecolumn
\begin{appendix}
\section{Posterior probability distributions of disc parameters}
\label{sec:pos}
\begin{figure*}[!ht]
	\includegraphics[width=0.9\textwidth]{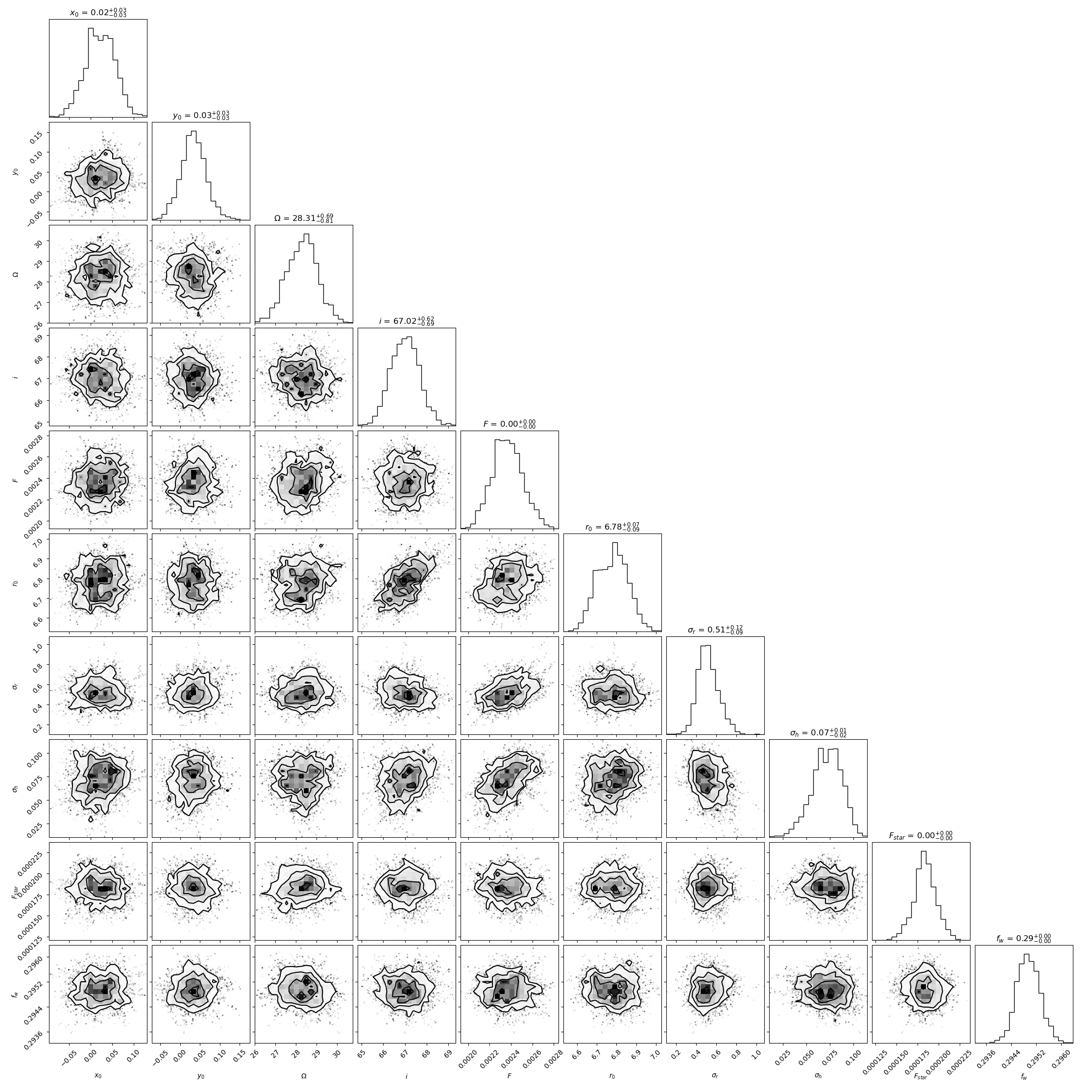}
	\caption{Corner plot showing the posterior probability distributions for the emcee runs used to identify the maximum amplitude probability model and calculate the uncertainties. The posteriors are mono-modal and well behaved except for the inclination, which runs up against the edge of the allowed parameter space (90$\degree$). }
	\label{fig:corner_g}
\end{figure*}

\begin{figure*}[!ht]
	\includegraphics[width=0.9\textwidth]{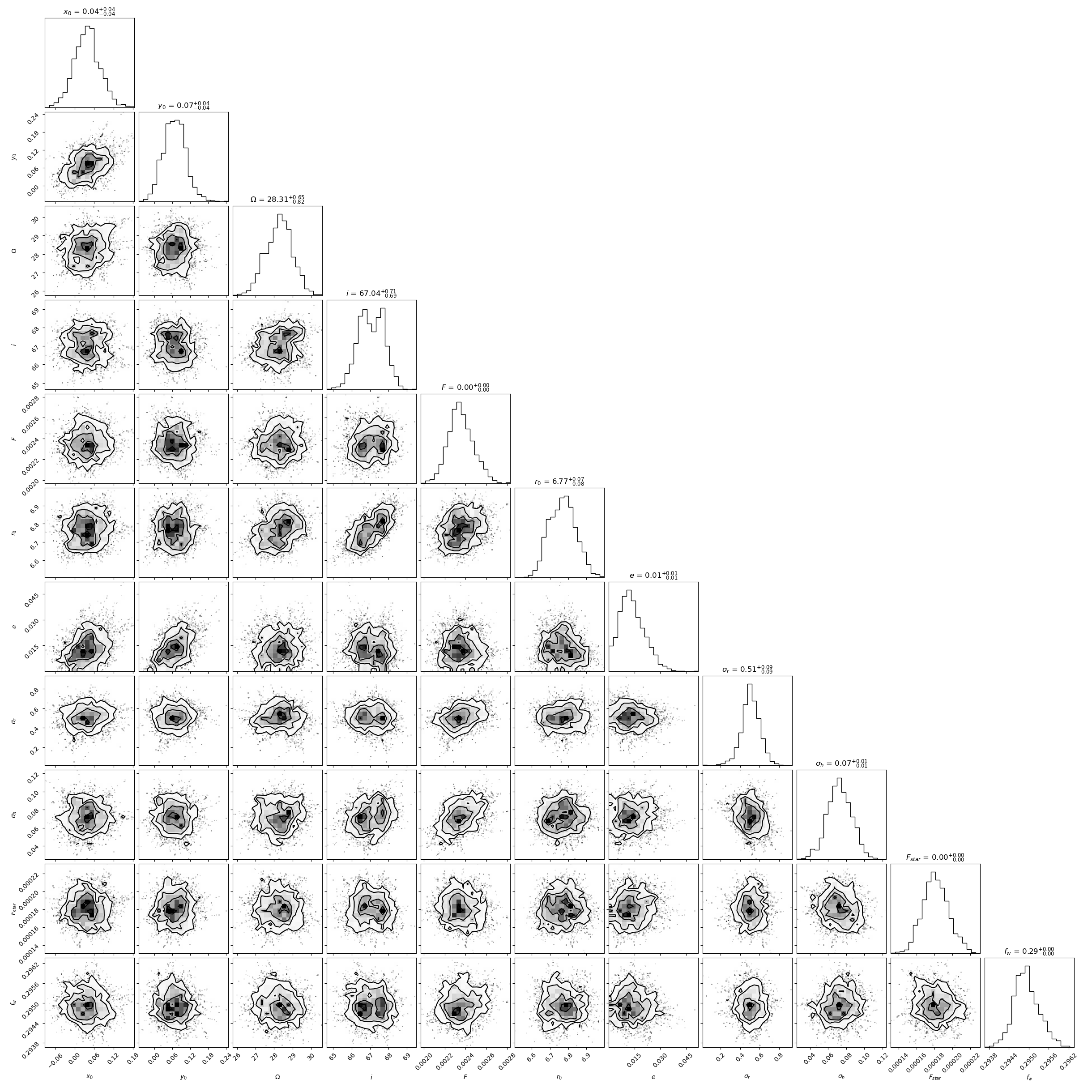}
	\caption{Corner plot showing the posterior probability distributions for the emcee runs used to identify the maximum amplitude probability model and calculate the uncertainties. The posteriors are mono-modal and well behaved except for the inclination, which runs up against the edge of the allowed parameter space (90$\degree$). }
	\label{fig:corner_e}
\end{figure*}

Figure~\ref{fig:corner_g} displays the posterior probability distributions of the disc parameters for the Gaussian model, while Fig.~\ref{fig:corner_e} presents those for the eccentric model. The distributions show that the parameters are well constrained and show little degeneracy.

\end{appendix}

\end{document}